\documentclass[letterpaper,10pt]{IEEEtran}
\IEEEoverridecommandlockouts
\usepackage{cite}
\usepackage{amsmath,amssymb,amsfonts}
\usepackage{algorithmic}
\usepackage{graphicx}
\usepackage{textcomp}
\usepackage{xcolor}

\usepackage[ruled,linesnumbered]{algorithm2e}
\usepackage{subcaption}
\usepackage{graphicx}
\usepackage{multirow}
\usepackage{upgreek}
\usepackage{hyperref}
\usepackage{fontawesome}
\usepackage{float}
\usepackage{pifont}
\usepackage{enumitem}
\usepackage{color}
\usepackage{array}
\usepackage{threeparttable}
\usepackage{cleveref}
\usepackage{graphbox}
\usepackage{booktabs}
\usepackage{siunitx}
\usepackage[switch]{lineno}
\usepackage[misc]{ifsym}

\crefname{figure}{Fig.}{Fig.}
\crefname{table}{TABLE}{TABLE}

\def\BibTeX{{\rm B\kern-.05em{\sc i\kern-.025em b}\kern-.08em
    T\kern-.1667em\lower.7ex\hbox{E}\kern-.125emX}}
\begin{document}

\title{BP-Im2col: Implicit Im2col Supporting AI Backpropagation on Systolic Arrays*\thanks{*: This work was supported by National Nature Science Foundation of China under NSFC No. 61802420 and 62002366.}}

\author{
    \IEEEauthorblockN{Jianchao Yang, Mei Wen$^{\textrm{\Letter}}$\thanks{${\textrm{\Letter}}$ : Corresponding author.}, Junzhong Shen, Yasong Cao, Minjin Tang, Renyu Yang, Jiawei Fei, Chunyuan Zhang}
    \IEEEauthorblockA{School of Computer Science and Technology, National University of Defense Technology, Changsha, China}
    \IEEEauthorblockA{\{yangjianchao16,meiwen,shenjunzhong,caoyasong,tangminjin14,yangrenyu,feijiawei11,cyzhang\}@nudt.edu.cn}\vspace{-2em}
}

\maketitle

\pagestyle{empty}  
\thispagestyle{empty}

\begin{abstract}
State-of-the-art systolic array-based accelerators adopt the traditional im2col algorithm to
accelerate the inference of convolutional layers.
However, traditional im2col cannot efficiently support AI backpropagation.
Backpropagation in convolutional layers involves performing transposed convolution and dilated
convolution, which usually introduces plenty of zero-spaces into the feature map or kernel.
The zero-space data reorganization interfere with the continuity of training and
incur additional and non-negligible overhead in terms of off- and on-chip storage, access and performance.
Since countermeasures for backpropagation are rarely proposed, we propose BP-im2col,
a novel im2col algorithm for AI backpropagation, and implement it in RTL on a TPU-like
accelerator.
Experiments on TPU-like accelerator indicate that BP-im2col reduces the backpropagation
runtime by 34.9\% on average, and reduces the bandwidth of off-chip memory and on-chip buffers by at
least 22.7\% and 70.6\% respectively, over a baseline accelerator adopting the traditional im2col.
It further reduces the additional storage overhead in the backpropagation process by at least 74.78\%.
\end{abstract}
\begin{IEEEkeywords}
im2col, AI backpropagation, systolic array
\end{IEEEkeywords}

\vspace{-0.3cm}
\section{Introduction}

State-of-the-art neural network accelerators adopt systolic arrays \cite{SystolicArray} to accelerate
the inference and training of convolutional neural networks (CNNs)
\cite{SIGMA,TendorDash,TPU,Procrustes}.
The existing systolic array-based accelerators largely adopt the
traditional im2col algorithm \cite{im2col} to lower the inference of convolutional layers
to general matrix multiplication (GEMM).
Backpropagation in convolutional layers involves performing more complicated transposed convolution
and dilated convolution, which is necessary to perform zero-insertions and
zero-paddings (collectively referred to as zero-spaces)
for the feature map or kernel.
According to our analysis, for convolutional layers with $stride \geq 2$,
the zero-spaces cause the sparsity of the lowered matrix to be as high as
about 75\%.

Existing accelerators \cite{SIGMA,TendorDash,TPU,Procrustes} use the same systolic
array-based platforms to speed up the inference and training of convolutional layers.
The core idea of solving zero-space of the input or kernel on systolic
array-based platforms is to pre-process
them to be zero-inserted and zero-padded in advance \cite{FlexiGAN}.
However, the data reorganization requires large amounts of memory access and interferes with the continuity
of training.
Even though part of the latency of data reorganization can be hidden in the training
process as a whole, it nevertheless increases the complexity of hardware control.
The transmission of zero-spaces also leads to very high
bandwidth requirements, which is more obvious for processors with
mismatched bandwidth and computing power.
Therefore, it is essential for the im2col algorithm to integrate zero-skipping mechanism.
Besides, explicit im2col
generates and stores a matrix-like copy of the input and kernel to facilitate further
matrix multiplication by PEs, which also incurs significant
performance and memory overhead for the convolution itself. This disadvantage can be avoided through the use of the
implicit im2col.

While numerous publicly available methods \cite{IndirectConvolution,ChannelLastIm2col,AvoidZeroSpacing}
describing the im2col algorithm only support the inference
of convolutional layers, countermeasures
for the backpropagation are rarely proposed.
Our contributions are summarized as follows:
\begin{itemize}[leftmargin=10pt]
    \item We propose a novel implicit im2col algorithm, named BP-im2col, which completely
          eliminates the zero-space data reorganization during backpropagation;
    \item We design and implement a TPU-like accelerator, integrated with the hardware
          implementation of BP-im2col.
          The address generation modules achieve low-overhead
          Non-Zero detection and avoid data reorganization during training;
    \item The proposed TPU-like accelerator reduces the backpropagation runtime by 34.9\% on average, and reduces the bandwidth
          of off-chip memory and on-chip buffers by at least 22.7\% and 70.6\% respectively,
          over a baseline accelerator adopting the traditional im2col. It also reduces the additional
          storage overhead in the backpropagation process.
\end{itemize}

For clarity, \Cref{tab:symbols} shows the meaning of the symbols used in this article.

\vspace{-0.2cm}
\section{Backpropagation of CNN}\label{section:Motivation}

The backpropagation involves calculating the loss of the input and the gradient of the
kernel.
\Cref{eq:backeq} outlines the training process \cite{TendorDash,SIGMA}.
After expressing the convolution as a matrix
multiplication ($Y=A\times B$) via im2col \cite{im2col},
the huge benefit of the very regular memory access
pattern produces a high ratio of floating-point operations per byte of data transferred.

\begin{small}\vspace{-0.3cm}
  \begin{equation}\label{eq:backeq}
    \setlength{\abovedisplayskip}{-6cm}
    \setlength{\belowdisplayskip}{-6cm}
    \begin{aligned}
      inference: &\quad I^{l+1} = I^l_{e} \ast W^l\\
      loss: &\quad \delta I^l =  \delta I^{l+1}_{ei} \ast Tr(rot_{180^{\circ}}{W^l})\\
      gradient: &\quad Tr(\delta W^l) = Tr(I^l_{e}) \ast Tr(\delta I^{l+1}_{i})
    \end{aligned}\vspace{-0.3cm}
  \end{equation}
\end{small}

{\setlength\tabcolsep{1pt}
\begin{table}[t]
    \setlength{\abovecaptionskip}{3pt}
    \setlength{\belowcaptionskip}{-10pt}
	\centering
    \begin{footnotesize}
	\caption{Meaning of symbols.}\label{tab:symbols}
	\begin{tabular}{cl}
		\toprule
		Symbol & Meaning \\
		\midrule
		$I^l, W^l$ & Input and kernel of the $l$-th convolutional layer. \\
		$\delta I^l, \delta W^l$ & The loss of $I^l$ and the gradient of $W^l$. \\
		$\ast,rot_{180^{\circ}}$ & Convolution symbol, $180^{\circ}$ kernel-wise rotation. \\
		$Tr(\cdot)$ &  Transpose the first two dimensions of $4d$ tensors. \\
		$e/i/ei$ & Zero-paddings, zero-insertions, and both. \\
		$B,C,H_i,W_i$ & Batch size, input channel, height and width of $I^l$. \\
        $N,K_h,K_w$ & Output channel, height and width of $W^l$. \\
        $H_o,W_o$ & Height and width of the output $I^{l+1}$. \\
        $S,P_h,P_w$& Stride, padding in height and width directions. \\
        $H_o^{\prime\prime}$& $H_o+(H_o-1)\cdot(S-1)$. \\
        $W_o^{\prime\prime}$& $W_o+(W_o-1)\cdot(S-1)$. \\
        $H_o^{\prime\prime\prime}$& $H_o+2(K_h-1-P_h)+(H_o-1)\cdot(S-1)$. \\
        $W_o^{\prime\prime\prime}$& $W_o+2(K_w-1-P_w)+(W_o-1)\cdot(S-1)$. \\
        \bottomrule
	\end{tabular}\vspace{-0.1cm}
    \end{footnotesize}
\end{table}
}

\subsubsection{\textbf{Loss calculation}}
\begin{figure}[t]\vspace{-0.3cm}
  \setlength{\abovecaptionskip}{-3pt}
  \setlength{\belowcaptionskip}{-5pt}
  \centering
      \includegraphics[width=0.75\linewidth]{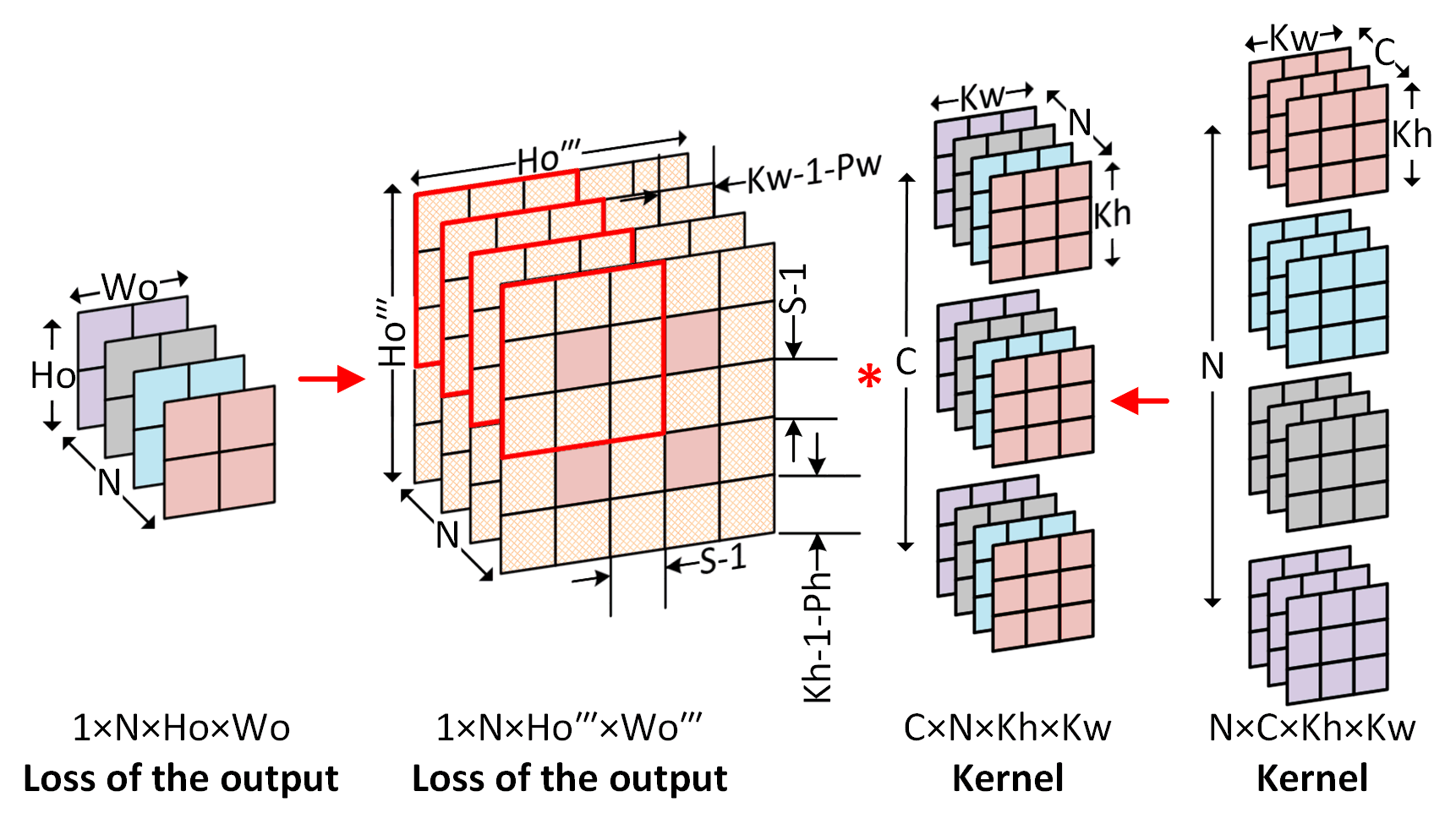}
  \caption{Loss calculation of convolutional layers.}
  \label{fig:Loss}\vspace{-0.3cm}
\end{figure}

\begin{figure}[t]\vspace{-0.3cm}
  \setlength{\abovecaptionskip}{-3pt}
  \setlength{\belowcaptionskip}{-3pt}
  \centering
  \begin{subfigure}{0.5\textwidth}
      \centering
      \includegraphics[width=\linewidth]{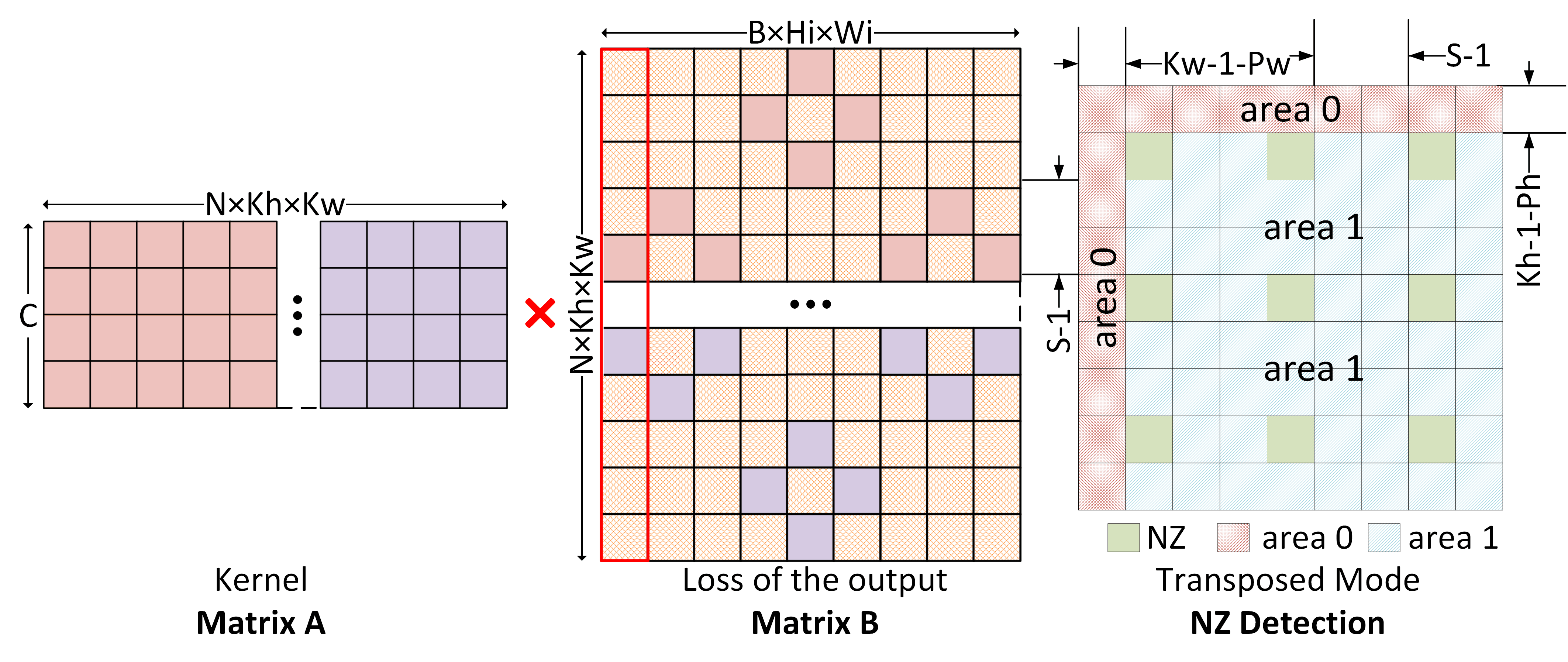}
  \end{subfigure}
  \caption{Traditional im2col of loss calculation and NZ detection of transposed mode. The data of matrix B marked by the red boxes in
           \Cref{fig:Loss} is expanded into the column marked by the red box in \Cref{fig:LossAll}.}
  \label{fig:LossAll}\vspace{-0.3cm}
\end{figure}

The difference between loss calculation and inference is
that loss calculation is realized by performing transposed convolution on the loss of the
output by the convolving kernel (see \Cref{eq:backeq}).
Another important difference is that the stride of the transposed convolution is a fixed value
of $1$. The transposed convolution and the im2col process of loss calculation are illustrated
in \Cref{fig:Loss} and \Cref{fig:LossAll}.
It can be observed that zero-insertions and zero-paddings of the loss of the output result in
more zero pixels in the convoluted feature map.
The combination of zero-paddings and zero-insertions introduces a huge amount of zero pixels to
matrix $B$ after im2col, and the ratio of zero pixels is
as high as 75\% to 93.91\% for popular convolutional neural networks.

\subsubsection{\textbf{Gradient Calculation}}
\begin{figure}[t]
  \setlength{\abovecaptionskip}{3pt}
  \setlength{\belowcaptionskip}{-3pt}
  \centering
      \includegraphics[width=0.75\linewidth]{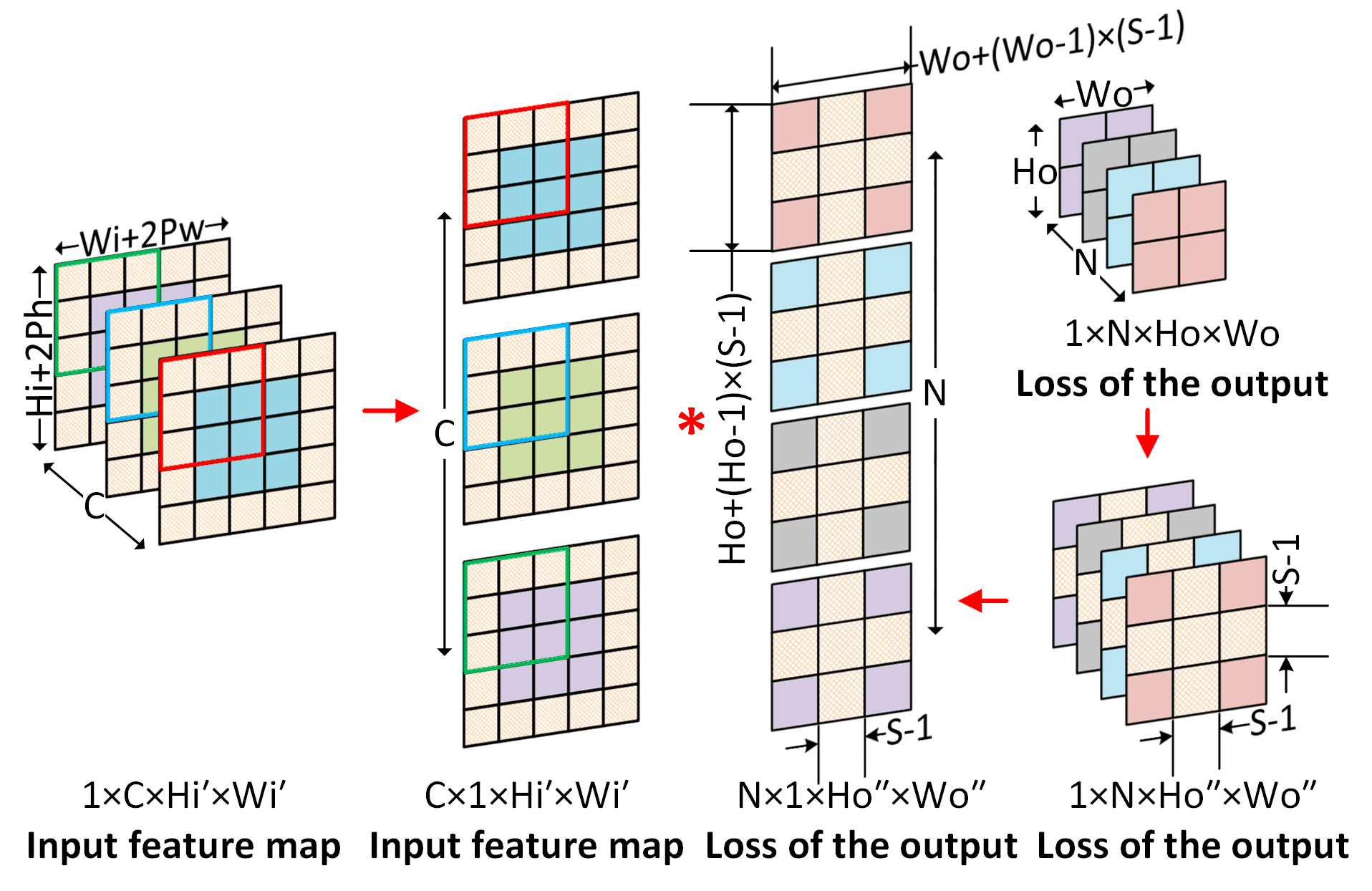}
  \caption{Gradient calculation of convolutional layers.}
  \label{fig:Gradient}\vspace{-0.4cm}
\end{figure}

\begin{figure}[t]
  \setlength{\abovecaptionskip}{-3pt}
  \setlength{\belowcaptionskip}{-3pt}
  \centering
  \begin{subfigure}{0.51\textwidth}
      \centering
      \includegraphics[width=\linewidth]{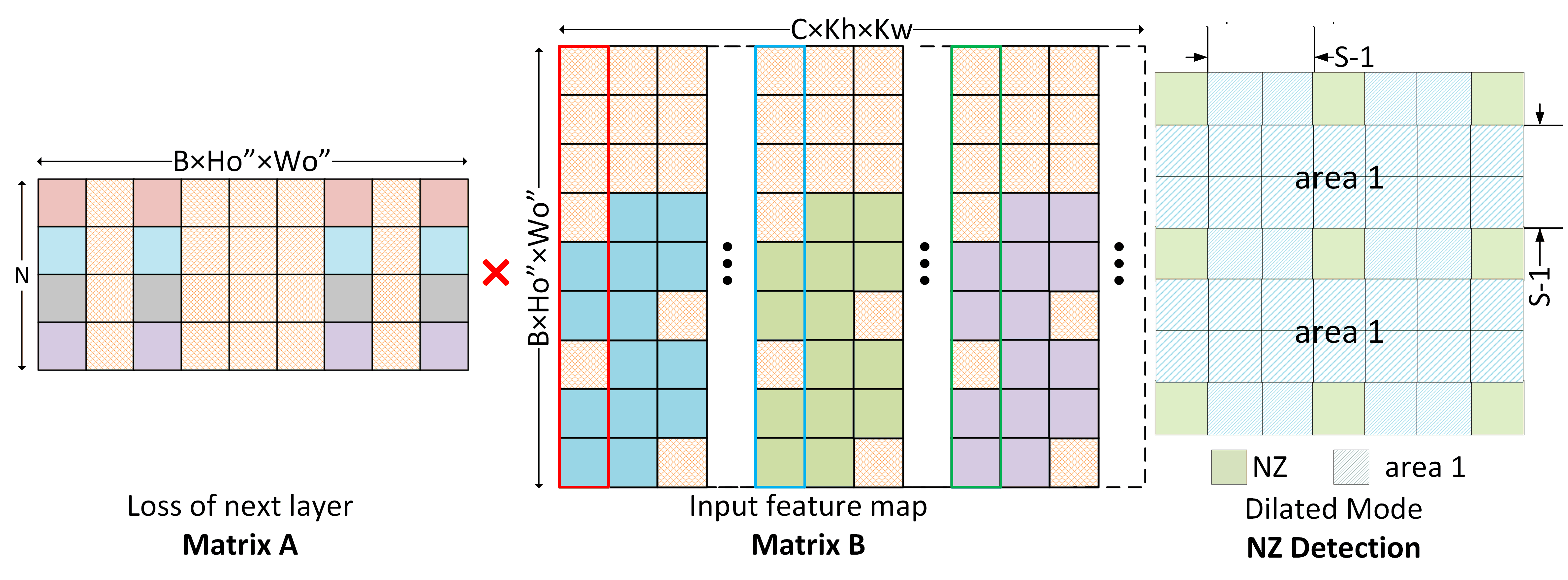}
  \end{subfigure}
  \caption{Traditional im2col of gradient calculation and NZ detection of dilated mode. The data of matrix B marked by the colored boxes in \Cref{fig:Gradient} is expanded into the column marked by the colored box in \Cref{fig:GradientAll}.}
  \label{fig:GradientAll}
\end{figure}

The gradient calculation is realized by
performing dilated convolution on the reorganized input by the reorganized loss of the
output (see \Cref{eq:backeq}).
As with the loss calculation, the stride of the dilated convolution
is a fixed value of $1$. We detail the reorganized steps of the input and the loss
of the output in \Cref{fig:Gradient}, while \Cref{fig:GradientAll} illustrates the im2col
process for gradient calculation.
The number of zeros introduced by the zero-padding of the input is
roughly the same as that introduced by the inference.
What causes the overall plenty of zeros is the zero-insertions for the loss of the output.
The zero pixels caused by zero-insertions for the loss of the output is extremely large,
and the ratio of zero pixels is as high as 74.8\% to 93.6\% for popular convolutional neural networks.

\vspace{-0.2cm}
\section{Algorithm and Hardware Design}\label{section:mapOnSA}

\subsection{Address Generation of BP-Im2col}\label{section:AddressGeneration}


When performing BP-im2col for loss calculation, we maintain a virtual matrix $B$
along with a virtual four-dimensional convoluted feature map with zero-spaces. We map the
addresses of virtual matrix $B$ to the virtual four-dimensional convoluted feature map with
zero-spaces, and then map it to the four-dimensional convoluted feature
map without zero-spaces, which is actually stored in the on-chip buffer.
For gradient calculation, the mapping of matrix $A$ with zero-spaces is similar to matrix $B$,
except that it does not need to perform im2col and has only zero-insertions.
\Cref{fig:addrmap} describes the address mapping of matrix $A$ and matrix $B$.

\vspace{-0.4cm}
\subsection{NZ Detection}\label{section:NZDection}

\subsubsection{\textbf{Transposed convolution mode}}\label{subsection:transposedmode}
For loss calculation, we divide the zero pixels
in a single channel into two areas: namely, one is composed of upper and left zero-paddings
(area $0$), while the other is composed of other zero-spaces (area $1$), which is shown in \Cref{fig:LossAll}.
The condition that a pixel $(h,w)$ is in area $0$ is: 
\begin{equation}\label{eq:condition1}\small\vspace{-0.2cm}
    h < K_h-1-P_h\ or\ w < K_w-1-P_w.\vspace{-0.1cm}
\end{equation}
Moreover, the condition that the pixel is in area $1$ is:
\begin{equation}\label{eq:condition2}\small\vspace{-0.1cm}
    [h-(K_h-1-P_h)] \%S > 0\ or\ [w-(K_w-1-P_w)]\%S > 0.\vspace{-0.1cm}
\end{equation}
We present the address mapping algorithm of matrix $B$ lowered during loss calculation in
\Cref{alg:Loss}.

\begin{algorithm}[t]\small
  \setlength{\abovecaptionskip}{0pt}
  \setlength{\belowcaptionskip}{-5pt}
  \caption{BP-im2col of transposed mode.}
  \label{alg:Loss}
    \KwIn{Address of a pixel in virtual matrix $B$, $addr_{in}$;}
    \KwOut{Address in the original feature map without zero-spaces, $addr_{out}$;}
    $row$, $col$ = $\lfloor addr_{in}/(B\cdot H_i\cdot W_i)\rfloor$, $addr_{in}\%(B\cdot H_i\cdot W_i)$;\\
    \label{alg:Loss:rowcol}
    $b$, $temp1$, $w_k$ = $\lfloor col/(H_i\cdot W_i)\rfloor$, $\lfloor row/K_w\rfloor$, $row\%K_w$;\\
    \label{alg:Loss:bw}
    $n$, $h_k$, $temp2$ = $\lfloor temp1/K_h\rfloor$, $temp1\%K_h$, $col\%(H_i\cdot W_i)$;\\
    $h, w$ = $\lfloor temp2/Wi\rfloor+h_k$, $temp2\%Wi+w_k$;\\
    \label{alg:Loss:nh}
    \eIf{$(h,w)$ satisfy \Cref{eq:condition1} or \Cref{eq:condition2}}{
        $addr_{out}$ = $NULL$; //Zero-spaces.
        \label{alg:Loss:addrout1}
    }{
        $h^{'}$, $w^{'}$ = $(h-(K_h-1-P_h), w-(K_w-1-P_w))/S$;\\
        \label{alg:Loss:bnhwnew}
        $addr_{out}$ = $b\cdot N\cdot H_o\cdot W_o + n\cdot H_o\cdot W_o + h^{'}\cdot W_o + w^{'}$;\\
        \label{alg:Loss:addrout2}
    }\vspace{-0.1cm}
\end{algorithm}

\subsubsection{\textbf{Dilated convolution mode}}\label{subsection:dilatedmode}
Assuming that a certain pixel to be calculated
is mapped to the position of the virtual convolving kernel with zero-insertions as $(h,w)$, the position
of the pixel in the channel is shown in \Cref{fig:LossAll}. The condition that this pixel
to be located in the zero pixel area (area $1$) is:
\begin{equation}\label{eq:condition3}\small\vspace{-0.2cm}
    h\%S > 0\ or\ w\%S > 0.
\end{equation}
Moreover, its target position in the actually stored convolving kernel is $(h/S, w/S)$.
We present the address mapping algorithm of matrix $A$ lowered during gradient calculation
in \Cref{alg:Gradient}.

\begin{algorithm}[t]\small
  \caption{BP-im2col of dilated mode.}
  \label{alg:Gradient}
    \KwIn{Address of a pixel in virtual matrix $A$, $addr_{in}$;}
    \KwOut{Address in original loss of the output without zero-insertions, $addr_{out}$;}
    $n$, $col$ = $\lfloor addr_{in}/(B\cdot H_o^{\prime\prime}\cdot W_o^{\prime\prime})\rfloor$, $addr_{in}\%(B\cdot
    H_o^{\prime\prime}\cdot W_o^{\prime\prime})$;\\
    \label{alg:Gradient:ncol}
    $temp$, $w$ = $\lfloor col/W_o^{\prime\prime}\rfloor$, $col\%W_o^{\prime\prime}$;\\
    \label{alg:Gradient:w}
    $b$, $h$ = $\lfloor temp/H_o^{\prime\prime}\rfloor$, $temp\%H_o^{\prime\prime}$;\\
    \label{alg:Gradient:bh}
    \eIf{$(h,w)$ satisfy \Cref{eq:condition3}}{
        $addr_{out}$ = $NULL$; //Zero-insertions.
        \label{alg:Gradient:addrout1}
    }{
        $h^{'}$, $w^{'}$ = $(h, w)/S$;\\
        \label{alg:Gradient:bnhwnew}
        $addr_{out}$ = $b\cdot N\cdot H_o\cdot W_o + n\cdot H_o\cdot W_o + h^{'}\cdot W_o + w^{'}$;\\
        \label{alg:Gradient:addrout2}
    }\vspace{-0.1cm}
\end{algorithm}

\vspace{-0.4cm}
\subsection{Hardware Design}

\begin{figure}[t]\vspace{-0.3cm}
  \setlength{\abovecaptionskip}{0pt}
  \setlength{\belowcaptionskip}{-1pt}
  \centering
  \includegraphics[width=0.38475\textwidth,height=0.2375\textheight]{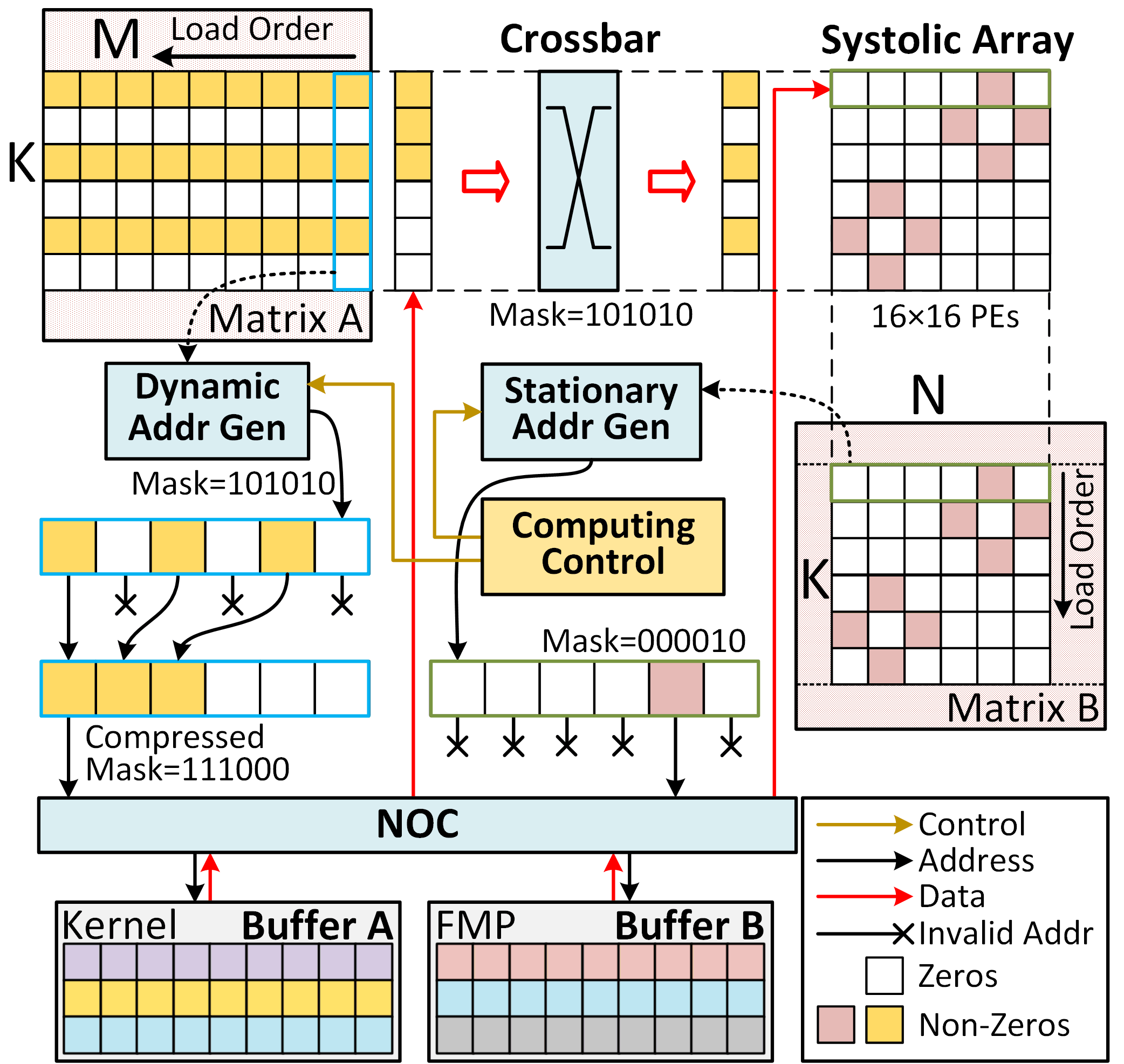}
  \caption{
  Address mapping and architecture of TPU-like.
  }\label{fig:addrmap}\vspace{-0.6cm}
\end{figure}

We implement a systolic array, named as TPU-like accelerator. It uses a $16\times 16$ systolic array as
the acceleration core and adopts the input-stationary data flow.
\Cref{fig:addrmap} illustrates
the architectural details. Both
buffer $A$ and buffer $B$ are double-buffered. Buffer $A$ supplies the data of
the dynamic lowered matrix $A$ for PEs, while buffer $B$ supplies that of the
stationary lowered matrix $B$.
We design $16$ FIFOs with different depths between
buffer $A$ and the systolic array to skew the data layout. To implement BP-im2col,
we use address generation and
compression logic to generate appropriate addresses for each block of matrix $A$ and matrix $B$,
and recover the data format for the compressed data that is transmitted back.


{
\setlength{\parindent}{0cm}
\textbf{\newline \textit{Transposed convolution mode.}}
}
In \Cref{fig:addrmap}, we describe
how a block of matrix $B$ is loaded onto the systolic array. The address generation module first generates
pixel addresses under the virtual stationary matrix $B$ view and we take $16$
channels to generate addresses in parallel during the address generation of matrix $B$ to
supply data
for $16$ PEs in each row of the systolic array.
We detect each address according to \Cref{section:NZDection} to filter
the zero pixels, and perform address mapping to generate the compressed
address of the actually stored feature map. After the data is
transmitted back, we send it directly to the PEs, according to its compressed mask.
When the data enters the systolic array, the zero pixel position
identified by the compressed mask is temporarily filled with zeros.

\vspace{-0.4cm}
{
\setlength{\parindent}{0cm}
\textbf{\newline \textit{Dilated convolution mode.}}
}
\Cref{fig:addrmap} also describes how a block of lowered matrix $A$ is loaded into the
systolic array.
The dynamic matrix address generation module generates addresses under the virtual dynamic matrix $A$ view.
The addresses of the dynamic matrix $A$ are continuous;
thus, we only generate the first address of the data in each row of blocks of matrix $A$ ($addr$), and the addresses of
the $16$ elements in this row are: $addr$, $addr+1$, $\cdots$, $addr+15$.
However, $16$ elements of a row block of matrix $A$ are not strictly
continuously stored for dilated convolution,
for the reason that there may be zeros that are not actually stored.
We therefore need all addresses of the $16$ channels to perform address mapping and NZ detection
to determine the non-zero position of the row elements.
Although the mapped addresses of the $16$ elements
in a row of matrix $A$ are not strictly consecutive, the non-zero elements are stored consecutively
in buffer $A$. We compress the non-consecutive $16$ mapped element addresses and
send only the address of the first non-zero element to buffer $A$.
The data transmitted back by
buffer $A$ is a continuous number of elements starting from the first non-zero element.
Then we recover the data arrangement through a crossbar
according to the original mask. Similarly, only the compressed addresses and non-zero data are
passed on to the chip.

\vspace{-0.3cm}
\section{Evaluation}
\textbf{TPU-like Experiment Setup.}
We implement the traditional im2col \cite{im2col} and BP-im2col on TPU-like accelerator. 
Our evaluation uses the
$FP32$ data type and a batch size of $2$.
The synthesis uses $ASAP7$, a 7 nm
predictive PDK library \cite{ASAP7}.

\textbf{Workload.}
We evaluate all convolutional layers with stride $\geq 2$ from several CNNs.
The \textit{"Original"} legend in figures is referred to the adoption of traditional im2col integrated with zero-space reorganization,
while the \textit{"Ours"} legend refers to the adoption of implicit BP-Im2col.


\vspace{-0.45cm}
\subsection{Overall Calculation Time}

\begin{figure}[t]\vspace{-0.3cm}
  \setlength{\abovecaptionskip}{0.2cm}
  \setlength{\belowcaptionskip}{-0.2cm}
  \centering
    \begin{subfigure}{0.235\textwidth}
      \centering
      \includegraphics[width=\linewidth]{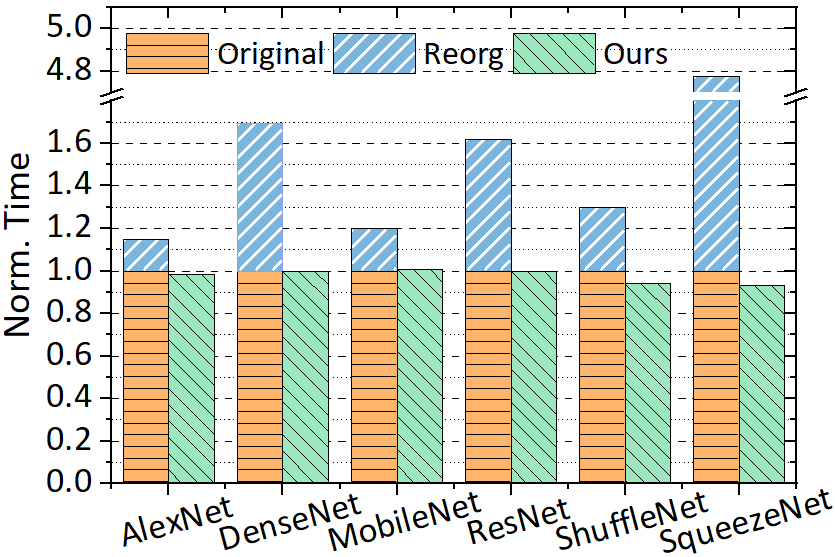}
        \caption{Loss calculation.}
        \label{fig:losstotaltime}
    \end{subfigure} 
    \begin{subfigure}{0.234\textwidth}
      \centering
      \includegraphics[width=\linewidth]{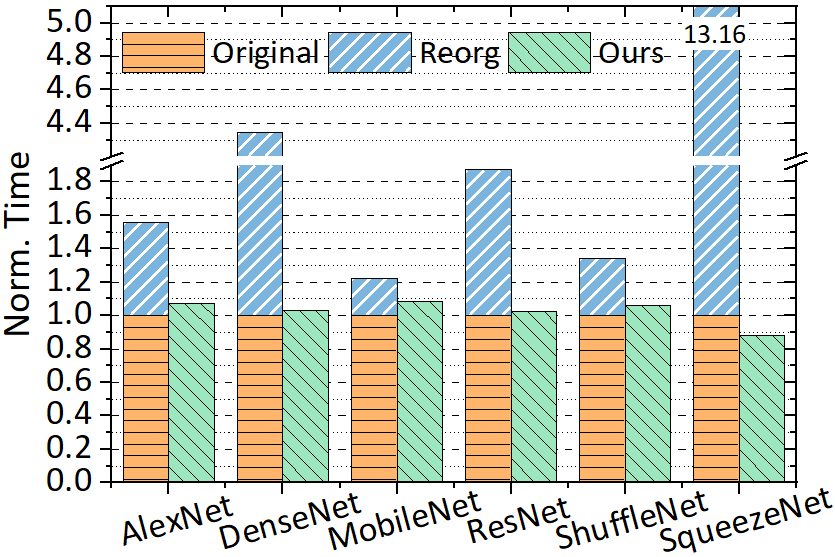}
        \caption{Gradient calculation.}
        \label{fig:gradtotaltime}
    \end{subfigure}
\caption{
\label{fig:tpuliketotaltime}
Performance comparison.
}\vspace{-0.3cm}
\end{figure}

We recorded the performing time of loss calculation
and gradient calculation during backpropagation.
\autoref{fig:losstotaltime} demonstrates that BP-im2col significantly reduces the loss calculation time
by 14.5\%, 41.2\%, 16.0\%, 38.3\%, 22.8\% and 79.0\% respectively, and that most of this gap stems
from the data reorganization of zero-spaces.
\autoref{fig:gradtotaltime} demonstrates that the gradient calculation time of BP-im2col is reduced by 31.3\%,
76.3\%, 17.7\%, 45.3\%, 20.9\% and 92.4\% respectively. BP-im2col greatly reduces the performance overhead
of loss calculation and gradient calculation caused by data reorganization.
\autoref{tab:layertime} also shows the runtime of loss calculation and gradient calculation of several
convolutional layers.

\begin{tiny}
\begin{table*}[!htbp]\vspace{-0.3cm}
\centering
\caption{Runtime of loss calculation and gradient calculation of several convolutional layers.}
\label{tab:layertime}
\begin{tabular}{|c|c|c|c|c|c|c|c|c|}
\hline



Convolution layers & \multicolumn{4}{c|}{Loss Calculation (cycles)} & \multicolumn{4}{c|}{Grad Calculation (cycles)} \\
\cline{2-9}

$H_i(W_i)/C/N/K_h(K_w)/$ & \multirow{2}{*}{BP-im2col} & \multicolumn{2}{c|}{Traditional im2col} & \multirow{2}{*}{Speedup} & \multirow{2}{*}{BP-im2col} & \multicolumn{2}{c|}{Traditional im2col} & \multirow{2}{*}{Speedup} \\
\cline{3-4}\cline{7-8}

$S/P_h(P_w)$ &  & Computation & Reorganization & & & Computation & Reorganization & \\
\cline{1-9}

224/3/64/3/2/0 & 8962102 & 8929989 & 37083360 & \textbf{5.13$\times$} & 2416476 & 2274645 & 37083360 & \textbf{16.29$\times$} \\
\cline{1-9}

112/64/64/3/2/1 & 10310400 & 10329856 & 3798997 & \textbf{1.37$\times$} & 9439744 & 8905216 & 3798997 & \textbf{1.35$\times$} \\
\cline{1-9}

56/256/512/1/2/0 & 9330688 & 9125888 & 15592964 & \textbf{2.65$\times$} & 11653120 & 11636736 & 15592964 & \textbf{2.34$\times$} \\
\cline{1-9}

28/244/244/3/2/1 & 8081314 & 8222247 & 1657646 & \textbf{1.22$\times$} & 8575509 & 8089919 & 1657646 & \textbf{1.14$\times$} \\
\cline{1-9}

14/1024/2048/1/2/0 & 11984896 & 11059200 & 6074461 & \textbf{1.42$\times$} & 15278080 & 15245312 & 6074461 & \textbf{1.40$\times$} \\
\hline
\end{tabular}\vspace{-0.3cm}
\end{table*}
\end{tiny}

\subsection{Off-chip Memory \& Buffer Bandwidth Occupation}

\begin{figure}[t]
  \setlength{\abovecaptionskip}{3pt}
  \setlength{\belowcaptionskip}{-1pt}
  \centering
    \begin{subfigure}{0.235\textwidth}
      \centering
      \includegraphics[width=\linewidth]{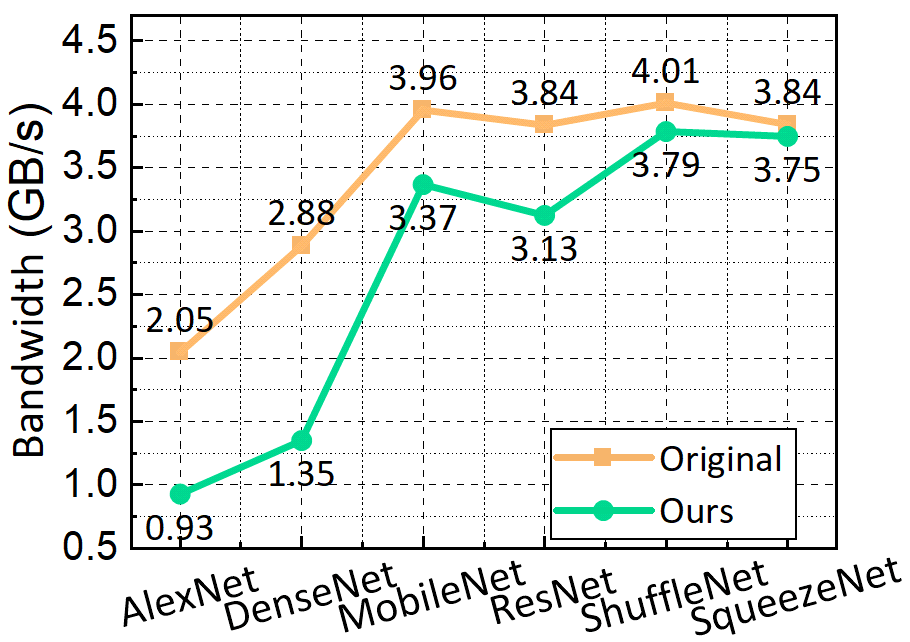}
        \caption{Loss calculation.}
        \label{fig:lossbbufferbw}
    \end{subfigure} 
    \begin{subfigure}{0.235\textwidth}
      \centering
      \includegraphics[width=\linewidth]{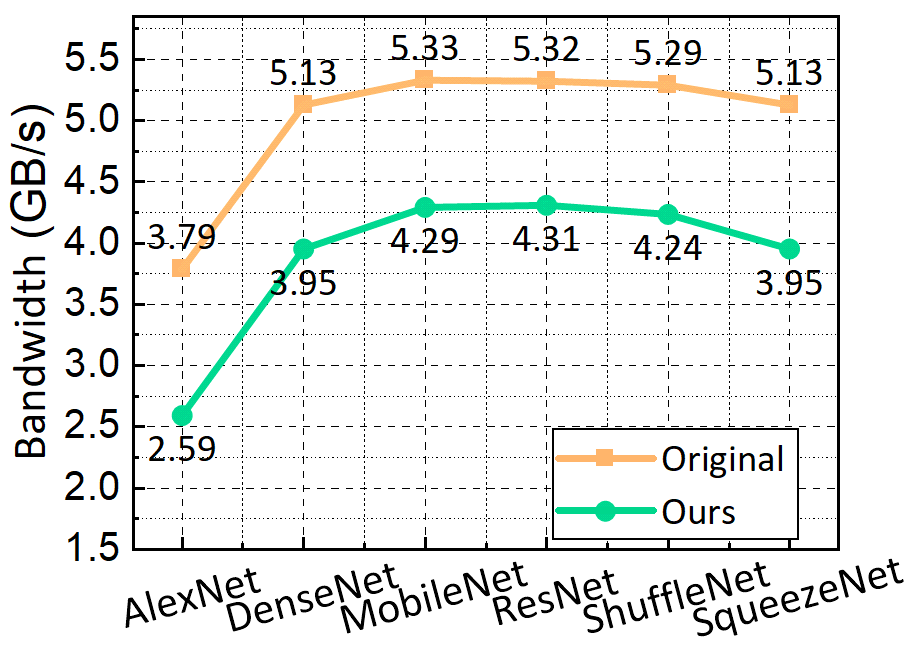}
        \caption{Gradient calculation.}
        \label{fig:gradabufferbw}
    \end{subfigure}
\caption{
\label{fig:bufferbw}
The bandwidth occupation of off-chip memory.
}\vspace{-0.6cm}
\end{figure}

\autoref{fig:lossbbufferbw} demonstrates that BP-im2col significantly
reduces the bandwidth occupation of data transmission to buffer $B$ during loss calculation:
specifically, it has a minimum reduction of 2.34\% (for SqueezeNet) and a maximum reduction of 54.63\% (for
AlexNet). \autoref{fig:gradabufferbw} further demonstrates that BP-im2col significantly
reduces the bandwidth occupation of data transmission to buffer $A$ during gradient calculation:
specifically, it has a minimum reduction of 18.98\% (for ResNet) and
a maximum reduction of 31.66\% (for AlexNet).

\begin{figure}[t]\vspace{-0.3cm}
  \setlength{\abovecaptionskip}{0pt}
  \setlength{\belowcaptionskip}{0pt}
  \centering
    \begin{subfigure}{0.235\textwidth}
      \centering
      \includegraphics[width=\linewidth]{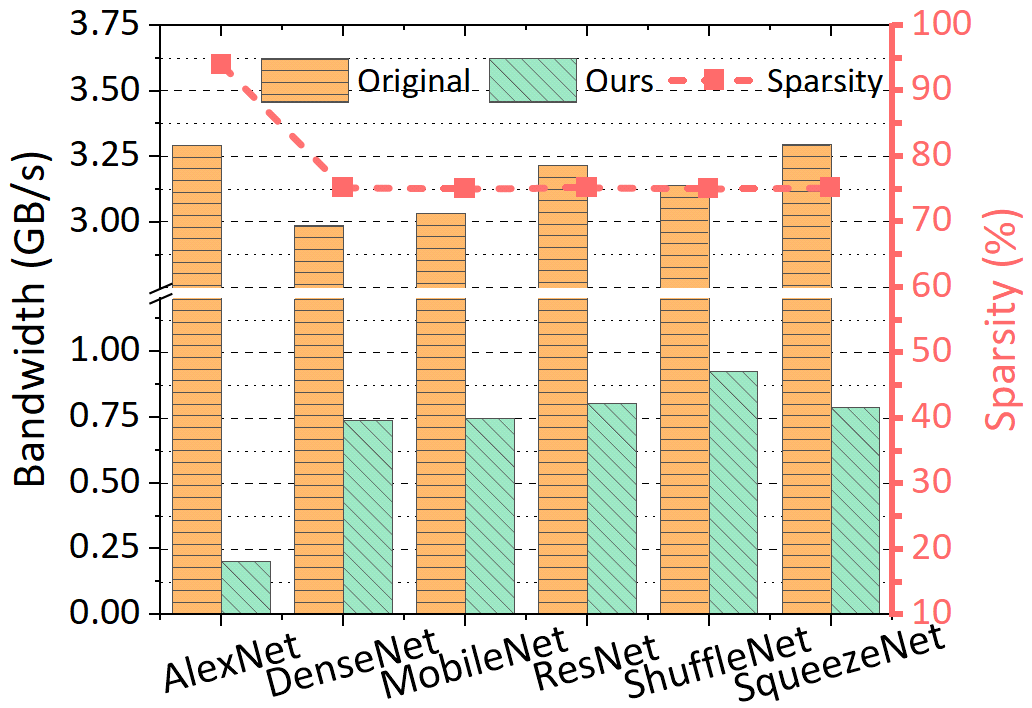}
        \caption{Bandwidth of buffer B during \hspace*{12pt}loss calculation.}
        \label{fig:lossbpebw}
    \end{subfigure} 
    \begin{subfigure}{0.235\textwidth}
      \centering
      \includegraphics[width=\linewidth]{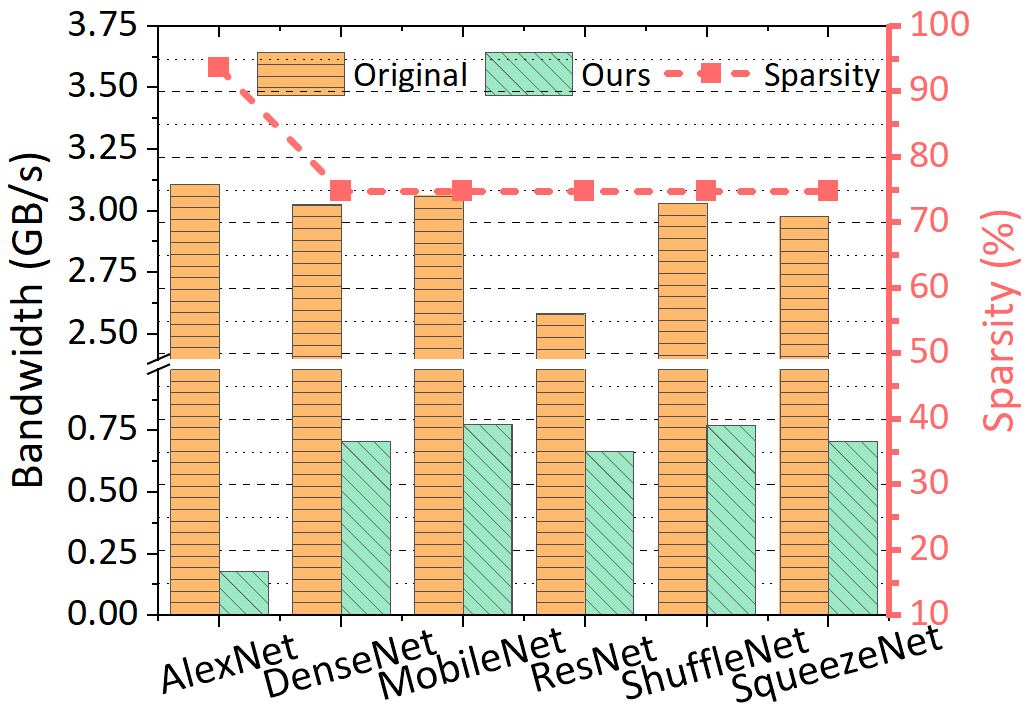}
        \caption{Bandwidth of buffer A during \hspace*{12pt}gradient calculation.}
        \label{fig:gradapebw}
    \end{subfigure}
\caption{
\label{fig:pebw}
The bandwidth occupation of on-chip buffers and the sparsity of calculation.
}\vspace{-0.6cm}
\end{figure}

\autoref{fig:lossbpebw} demonstrates that BP-im2col reduces the bandwidth occupation of buffer $B$ during loss calculation by 93.90\%, 75.36\%, 75.45\%, 75.04\%, 70.56\%, and 76.15\%, respectively. The ratio
of the bandwidth occupation reduction of buffer $B$ is close to the sparsity of the loss of
the output during loss calculation.
\autoref{fig:gradapebw} demonstrates that BP-im2col reduces the bandwidth occupation of buffer $A$ by 94.23\%, 76.67\%, 74.70\%,
74.15\%, 74.53\%, and 76.30\%, respectively, which is also close to the sparsity of the loss of
the output during gradient calculation.

\vspace{-0.3cm}
\subsection{Prologue Latency Overhead \& Area Overhead}

{\setlength\tabcolsep{3.5pt}
\begin{tiny}
\begin{table}[!htbp]
\setlength{\abovecaptionskip}{3pt}
\setlength{\belowcaptionskip}{0pt}
\centering
\caption{Prologue latency for two matrix
         address generation modules with sufficient network bandwidth.}
\begin{tabular}{ccccc}
\hline
\multirow{2}{*}{Module}&\multicolumn{2}{c}{Loss calculation}&\multicolumn{2}{c}{Gradient calculation}\\
\cline{2-3}\cline{4-5}
&Dynamic&Stationary&Dynamic&Stationary\\
\hline
Traditional im2col&0 cycle&51 cycles&0 cycle&51 cycles\\
BP-im2col&0 cycle&68 cycles&68 cycles&51 cycles\\
\hline
\end{tabular}
\label{tab:prologuelatency}
\end{table}
\end{tiny}
}
The prologue latency introduced by fixed-point dividers
from address mapping to completion of on-chip buffer address calculation, as shown in
\autoref{tab:prologuelatency}. And the area overhead of the address generation modules after adopting the traditional im2col
and BP-im2col in hardware is shown in \autoref{tab:areacomp}.

{\setlength\tabcolsep{3.5pt}
\begin{tiny}
\begin{table}[!htbp]
\setlength{\abovecaptionskip}{3pt}
\setlength{\belowcaptionskip}{0pt}
\centering
\caption{Area overhead of
         address generation modules.}
\begin{tabular}{ccccc}
\hline
\multirow{2}{*}{Module}&\multicolumn{2}{c}{Traditional im2col}&\multicolumn{2}{c}{BP-im2col}\\
\cline{2-5}
&Area $(\mu m^2)$&Ratio $(\%)$&Area $(\mu m^2)$&Ratio $(\%)$\\
\hline
Dynamic&5103&0.23&56628&2.44\\
Stationary&53268&2.42&121009&5.22\\
\hline
\end{tabular}
\label{tab:areacomp}\vspace{-0.3cm}
\end{table}
\end{tiny}
}

\section{Conclusion}

We propose an implicit im2col algorithm for AI backpropagation, named BP-im2col with the goal of better adapting the
training of convolutional layers mapping on systolic arrays. We
design and implement the hardware address generation modules based on the TPU-like accelerator, and further develop
special optimizations for the hardware based on the accelerator’s architectural characteristics.
However, our design does not support sparse computation at this stage, and the crossbar still occupy
a very large on-chip area after being pruned.
In the
future, we will further optimize sparse computation and data
flow for the computing modes of the TPU-like accelerator.

\bibliographystyle{IEEEtran}
\bibliography{IEEEabrv,IEEEexample.bib}

\end{document}